# Accelerating MR Imaging via Deep Chambolle-Pock Network*

Haifeng Wang, Jing Cheng, Sen Jia, Zhilang Qiu, Caiyun Shi, Lixian Zou, Shi Su, Yuchou Chang, Yanjie Zhu, Leslie Ying, and Dong Liang, *Senior Member, IEEE*

*Abstract*— Compressed sensing (CS) has been introduced to accelerate data acquisition in MR Imaging. However, CS-MRI methods suffer from detail loss with large acceleration and complicated parameter selection. To address the limitations of existing CS-MRI methods, a model-driven MR reconstruction is proposed that trains a deep network, named CP-net, which is derived from the Chambolle-Pock algorithm to reconstruct the in vivo MR images of human brains from highly undersampled complex k-space data acquired on different types of MR scanners. The proposed deep network can learn the proximal operator and parameters among the Chambolle-Pock algorithm. All of the experiments show that the proposed CP-net achieves more accurate MR reconstruction results, outperforming state-of-the-art methods across various quantitative metrics.

## I. Introduction

Accelerating MR imaging has been an ongoing research topic since its invention in the 1970s. Among variety of acceleration techniques, compressed sensing (CS) has become an important strategy using sub-Nyquist samples based on sparse prior during the past decades [1,2]. To recover the signals from sampled measurements, CS-based MR reconstruction can be formulated as an optimization problem which consists of the data consistency and the sparsity constraint. A number of iterative algorithms have been developed to solve the reconstruction problem, such as FISTA [3], ADMM [4], Chambolle-Pock algorithm [5], etc. Among them, Chambolle-Pock algorithm is a primal dual algorithm that solves the optimization problem simultaneously with its dual. And it has been successfully applied in medical image reconstruction due to its general coverage on many optimization formulations of interest [6]. Previously, the Chambolle-Pock algorithm has been applied by our group for accelerating MR imaging [7].

Although CS-based method can achieve high performance with many theoretical guarantees [1], it's challenging to determine the numerical uncertainties in the model such as the optimal sparse transformations, sparse regularizer in the transform domain, regularization parameters and the involve parameters of the optimization algorithm.

Recently, deep learning has demonstrated tremendous success and has become a growing trend in general data analysis [8-11]. It also has been applied successfully in medical imaging such as Ultrasound, CT, MRI, PET and so on [12-16]. Inspired by such success, deep learning has been applied in MR reconstruction and shown potential to significantly speed up MR acquisition and improve image quality [16-27]. Deep learning–based MR reconstruction can be roughly divided into two categories: data-driven [16-22] and model-driven [23-26]. Data-driven methods directly learn an end-to-end mapping between the input and the output with large training datasets and long training time. Model-driven networks formulate image reconstruction as unrolling the procedure of an iterative optimization algorithm to a network, while learning the transforms and parameters in the model.

In this work, we have proposed to incorporate the Chambolle-Pock algorithm into a network for accelerating MR imaging, which was named as CP-net, to learn the proximal operator and parameters among the Chambolle-Pock algorithm. This is inspired by the learned primal dual in CT reconstruction [14] and its application in MR reconstruction [27]. Whereas in our work, the network could handle complex data and maintain the algorithm structure of Chambolle-Pock algorithm to guarantee convergence. The proposed method can reconstruct images from highly undersampled complex k-space data for accelerating MR imaging. After many in vivo human brain experiments on different types of MR scanners, our results have shown that the proposed method can achieve superior results compared to traditional CS method [28] and the current state-of-the-art model-driven method ADMM-net [23] and data-driven method D5-C5 [21].

## II. Methods

In general, the problem to reconstruct the images from sub-Nyquist sampled k-space data can be written as

$$\min F(Ap) + G(p) \qquad (1)$$

where $F(Ap)$ is the data consistency term and can be expressed as $F(Ap) = \|Ap - y\|_2^2$, $A$ is the encoding matrix,

*Research supported in part by the National Natural Science Foundation of China (61871373, 81729003), Natural Science Foundation of Guangdong Province (2018A0303130132), Guangdong Provincial Key Laboratory of Medical Image Processing (2017A050501026), Guangdong Provincial Key Laboratory of Magnetic Resonance and Multimodality Imaging (2014B030301013), Shenzhen Key Laboratory for MRI (CXB201104220028A), Strategic Priority Research Program of Chinese Academy of Sciences (XDB25000000), and Shenzhen Key Laboratory of Ultrasound Imaging and Therapy (ZDSYS20180206180631473).

H Wang, J Cheng, S Jia, Z Qiu, C Shi, L Zou, S Su and Y Zhu are with Paul C. Lauterbur Research Centre for Biomedical Imaging, Shenzhen Institutes of Advanced Technology, Chinese Academy of Sciences, Shenzhen, Guangdong, China (e-mail: { hf.wang1, jing.cheng, sen.jia, zl.qiu, cy.shi, lx.zou, shi.su, yj.zhu}@siat.ac.cn)

Y Chang is with the Computer Science Department, University of Houston-Downtown, Houston, TX 77002 USA (e-mail: changy@uhd.edu).

L Ying is with the Biomedical Engineering and Electrical Engineering Departments, State University of New York, Buffalo, NY 14260 USA (e-mail: leiying@buffalo.edu).

D Liang is with Research Center and Medical AI, and Paul C. Lauterbur Research Centre for Biomedical Imaging, Shenzhen Institutes of Advanced Technology, Chinese Academy of Sciences, Shenzhen, Guangdong, China (corresponding author to provide phone: 86-0755-86392274; e-mail: dong.liang @ siat.ac.cn).

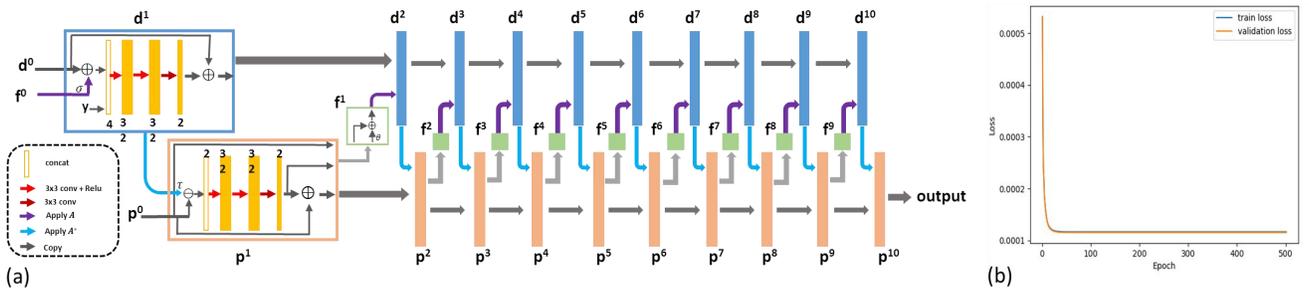

Figure 1. (a) Architecture of the proposed CP-net. The dual iterates are given on the first row, while the primal iterates are on the last row and the linking parts are in the green boxes. The architectures are illustrated in the corresponding enlarged boxes. (b) The plot of training (red line) and validating errors (blue line) versus epoch.

TABLE I. The detailed comparison on performance metrics of different reconstruction methods with different acceleration factor on the axial dataset.

| | R=4 | | | | | R=5 | | | | | R=6 | | | | |
|---|---|---|---|---|---|---|---|---|---|---|---|---|---|---|---|
| | CP-net | ADMM-net | D5-C5 | Rec_PF | Zero-filling | CP-net | ADMM-net | D5-C5 | Rec_PF | Zero-filling | CP-net | ADMM-net | D5-C5 | Rec_PF | Zero-filling |
| MSE | **1.59e-4** | 3.79e-4 | 3.58e-4 | 7.61e-4 | 0.0024 | **2.44e-4** | 6.69e-4 | 5.62e-4 | 0.0023 | 0.0070 | **2.54e-4** | 7.43e-4 | 5.92e-4 | 0.0017 | 0.0053 |
| SSIM | **0.9716** | 0.9293 | 0.9696 | 0.8912 | 0.7615 | **0.9581** | 0.8888 | 0.9532 | 0.7970 | 0.6270 | **0.9568** | 0.8809 | 0.9515 | 0.8090 | 0.6556 |
| PSNR | **35.0182** | 34.2052 | 34.4635 | 31.1818 | 26.1451 | **33.1428** | 31.7450 | 32.4989 | 26.3679 | 21.5505 | **32.9555** | 31.2864 | 32.2787 | 27.6516 | 22.7654 |

here $A$ denotes the undersampled Fourier encoding, $p$ is the image to be reconstructed, $y$ is the sampled k-space data, and $G(p)$ denotes the regularization term with respect to the image prior information.

With the Chambolle-Pock algorithm, the MR reconstruction problem (1) can be solved as follows:

$$\begin{cases} d_{n+1} = prox_{\sigma}[F^*](d_n + \sigma A\bar{p}_n) \\ p_{n+1} = prox_{\tau}[G](p_n - \tau A^* d_{n+1}) \\ \bar{p}_{n+1} = p_{n+1} + \theta(p_{n+1} - p_n) \end{cases} \quad (2)$$

where $d$ is the dual vector in k-space, $\sigma$, $\tau$ and $\theta$ are algorithm parameters. $F^*$ is the convex conjugate of the function $F$, $A$ is the corresponding undersampled Fourier operator and $prox$ denotes the proximal mapping which can work directly with non-smooth objective functions.

Here, a parameterized operator, where the parameters are learned by the network from training data, is used to replace the original proximal operator, thus the whole algorithm can be formed as:

$$\begin{cases} d_{n+1} = \Gamma(d_n + \sigma A\bar{p}_n, y) \\ p_{n+1} = \Lambda(p_n - \tau A^* d_{n+1}) \\ \bar{p}_{n+1} = p_{n+1} + \theta(p_{n+1} - p_n) \end{cases} \quad (3)$$

In the CP-net, the primal proximal $\Lambda$, dual proximal $\Gamma$, parameters $\sigma$, $\tau$ and $\theta$ would be learned from training data.

The structure of the proposed CP-net model is shown in Fig. 1(a). CP-net consists of three parts: dual iterates, primal iterates and a linking part. The dual and primal iterates have the same architecture with 3 convolutional layers in each box. To train the network more easily, we made it a residual network. As MR data is complex, the real and imaginary parts of the data are separately saved in two channels of the network. The convolutions involved are all 3×3 pixels in size, and the number of channels is 2-32-32-2 for each primal update, and 4-32-32-2 for the dual update. The original input undersampled k-space data is supplied to the dual iterates. The nonlinearities are chosen to be Rectified Linear Unites (ReLU) functions and we let the number of iterations be 10. Since each iteration involves 2 3-layer networks, the total depth of the CP-net is 60 convolutional layers.

We trained the network using in-vivo MR datasets. Overall 200 fully sampled multi-contrast data from 8 subjects with a 3T scanner (MAGNETOM Trio, SIEMENS AG, Erlgen, Germany) were collected and informed consent was obtained from the imaging object in compliance with the IRB policy. The fully sampled data was acquired by a 12-channel head coil with matrix size of 256×256 and combined to single-channel data and then retrospectively undersampled using Poisson disk sampling mask. After normalization and image augmentation, we got 1600 k-space data, where 1400 for training and 200 for validation. The model was trained and evaluated on an Ubuntu 16.04 LTS (64-bit) operating system equipped with a Tesla TITAN XP Graphics Processing Unit (GPU, 12GB memory) in the open framework Tensorflow with CUDA and CUDNN support. The mean square error was chosen as the loss function, and we trained the networks by minimizing the loss function using the ADAM optimizer in TensorFlow. The performance of training convergence is illustrated in Fig. 1(b).

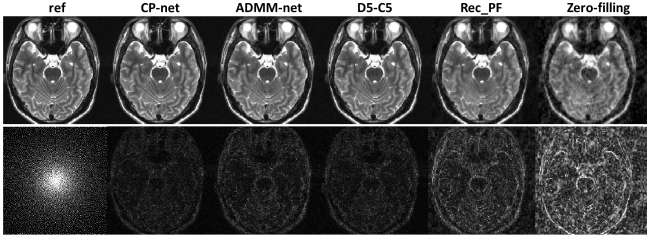

Figure 2. Comparison (Axial view) of the different reconstruction methods with R = 6. The images in second row are the sampling mask and the corresponding error maps between the reconstructions and the reference. The proposed achieves the highest accuracy reconstruction.

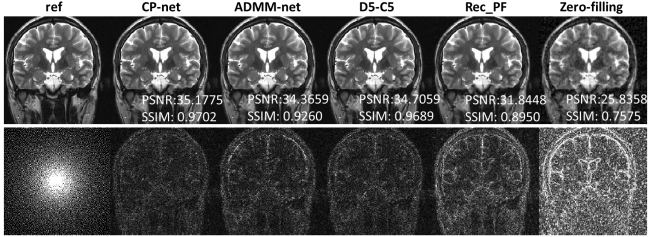

Figure 3. Comparison (Coronal view) of reconstruction methods with R=4. The proposed CP-net achieves better reconstruction than others.

Three human brain datasets with different view of sections from the SIEMENS 3T scanner were used as the test data and we also have tested the proposed model on the human data acquired from another 3T scanner (uMR 790, United Imaging Healthcare, Shanghai, China).

III. RESULTS

We used the data from three different views to evaluate the model and compared the CP-net with other CS-MRI reconstruction methods: 1) zero-filling, the inverse Fourier transform of the undersampled k-space data, 2) Rec_PF, traditional CSMR reconstruction method, 3) ADMM-net, state-of-the-art model-driven CSMR reconstruction method, 4) D5C5, an end-to-end deep learning method with data consistency. Several similarity metrics, including MSE, SSIM and PSNR, were used to compare the reconstruction results of different methods to reference image from fully sampled data.

From quantitative metrics shown in Table 1, on the axial dataset with different acceleration factors, the proposed CP-net all shows significant improvements over other methods. Fig. 2 illustrates the visual comparison and error distributions with acceleration factor of 6 on the axial data. It can be seen from Fig. 2 that the proposed method can obtain better image quality.

Figure 3 shows the visual comparison and error distributions from coronal view with a reduction factor of 4.

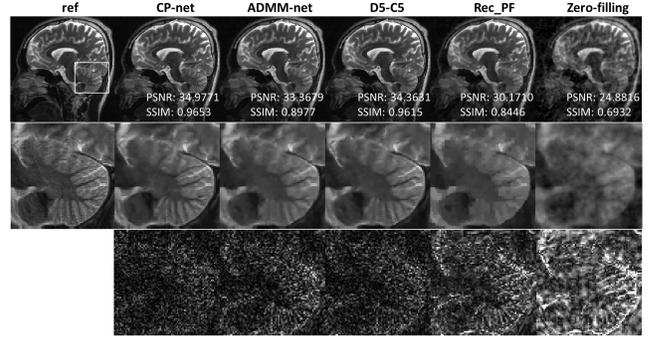

Figure 4. Visual comparisons (Sagittal view) for R=6. The enclosed part is enlarged for a close-up comparison. The zoom-in visualization and the corresponding error maps show the improvement of the proposed in detail preservation.

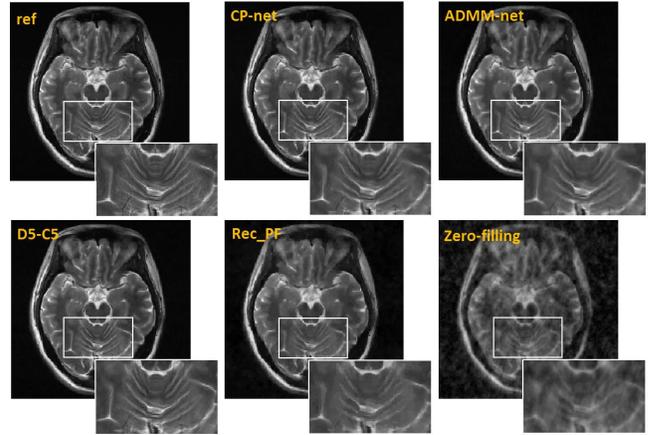

Figure 5. Comparison of reconstruction methods for R=5 with data from UIH scanner. The zoom-in visualization shows that the CP-net preserves more details.

Reconstruction accuracy is significantly improved using CP-net, which is indicated by both error maps and the quantitative metrics.

The reconstructions of different methods on sagittal data with R=6 are demonstrated on Fig. 4. The zoom-in views are on the second row and the corresponding error maps on the last row. It can be seen that CP-net preserves more fine details while removing the undersampling artifacts. Detailed accuracy metrics are also shown in figure.

Fig. 5 shows the reconstructions of data from UIH scanner, which of zoom-in visualization shows that the proposed method preserves more details via the trained CP-net. Therefore, the image results presented that the proposed method could robust overcome the traditional CS method and current ADMM-net method via image data from different types of MR scanners.

IV. CONCLUSION

In sum, we proposed an effective deep Chambolle-Pock network for accelerating MR imaging. The architecture of the proposed CP-net is defined over a data flow graph determined by the CP algorithm. Here, CP-net not only learns all the parameters in original algorithm but also learns the proximal operators for MR reconstruction. The current experimental results on in vivo human brain data have shown the highly efficient reconstruction of the proposed CP-net in terms of both artifacts removal and detail preservation. In the future, the proposed CP-net method will be applied those MRI applications of nonlinear gradients [29, 30].

## V. Acknowledgement

The authors will greatly thank Dr. Ozan Öktem for sharing their codes.

## References

[1] D. L. Donoho, "Compressed sensing," *IEEE Trans. Inf. Theory*, vol. 52, no. 4, pp. 1289–1306, 2006.
[2] M. Lustig, D. Donoho, and J. M. Pauly, "Sparse mri: The application of compressed sensing for rapid mr imaging," Magn. Reson. Med., vol. 58, no. 6, p. 11821195, 2007.
[3] A. Beck and M. Teboulle, "A fast iterative shrinkage-thresholding algorithm for linear inverse problems," *SIAM J. Img. Sci.*, vol. 2, no. 1, pp. 183–202, 2009.
[4] S. Boyd, N. Parikh, E. Chu, B. Peleato, and J. Eckstein, "Distributed optimization and statistical learning via the alternating direction method of multipliers," *Found. Trends Mach. Learn.*, vol. 3, no. 1, pp. 1–122, 2011.
[5] A. Chambolle and T. Pock, "A First-Order Primal-Dual Algorithm for Convex Problems with Applications to Imaging," *J. Math. Imaging Vis.,* vol. 40, no. 1, pp. 120-145, Mar 2011.
[6] E. Y. Sidky, J. H. Jorgensen, and X. C. Pan, "Convex optimization problem prototyping for image reconstruction in computed tomography with the Chambolle-Pock algorithm," *Physics in Medicine and Biology,* vol. 57, no. 10, pp. 3065-3091, 2012.
[7] J. Cheng, S. Jia, H. Wang, and D. Liang. "MR image reconstruction using the Chambolle-Pock algorithm." *In Proc. of Joint Annual Meeting ISMRM-ESMRMB*, Paris, France, no. 3552, 2018.
[8] G. E. Hinton, and R. R. Salakhutdinov. "Reducing the dimensionality of data with neural networks." *Science* 313, no. 5786 (2006): 504-507.
[9] A. Krizhevsky, I. Sutskever, and G. E. Hinton. "Image-net classification with deep convolutional neural networks." *In Advances in neural information processing systems*, pp. 1097-1105. 2012.
[10] M. I. Jordan, and T. M. Mitchell. "Machine learning: Trends, perspectives, and prospects." *Science* 349, no. 6245 (2015): 255-260.
[11] Y. LeCun, Y. Bengio, and G. E. Hinton. "Deep learning." *Nature* 521, no. 7553 (2015): 436.
[12] H. Chen, D. Ni, J. Qin, S. Li, X. Yang, T. Wang, P. A. Heng. "Standard Plane Localization in Fetal Ultrasound via Domain Transferred Deep Neural Networks." *IEEE J Biomed Health Inform*. vol. 19, no. 5, pp. 1627-36, Sep. 2015.
[13] H. Chen, Y. Zhang, W. Zhang, P. Liao, K. Li, J. Zhou, G. Wang. "Low-dose CT via convolutional neural network." *Biomed Opt Express.* vol.8, no. 2, pp. 679-694, Jan.. 2017.
[14] J. Adler, O. Oktem. "Learned Primal-Dual Reconstruction." *IEEE Trans Med Imaging.* vol. 37, no. 6, pp. 1322-1332, 2018.
[15] K. Gong, J. Guan, K. Kim, X. Zhang, J. Yang, Y. Seo, G. El Fakhri, J. Qi, Q. Li. "Iterative PET Image Reconstruction Using Convolutional Neural Network Representation." *IEEE Trans Med Imaging.* Sep. 2018. doi: 10.1109/TMI.2018.2869871.
[16] S Wang, Z Su, L Ying, et al. "Accelerating magnetic resonance imaging via deep learning," *ISBI2016*, pp. 514-517.
[17] K. Kwon, D. Kim, and H. Park, "A parallel MR imaging method using multilayer perceptron," *Medical Physics*, vol. 44, no. 12, pp. 6209–6224, 2017.
[18] Y. Han, J. Yoo, H. H. Kim, H. J. Shin, K. Sung, and J. C. Ye, "Deep learning with domain adaptation for accelerated projection-reconstruction MR," *Magn Reson Med*, vol. 80, no. 3, pp. 1189–1205, 2018.
[19] B. Zhu, J. Z. Liu, S. F. Cauley, B. R. Rosen, and M. S. Rosen, "Image reconstruction by domain-transform manifold learning," *Nature*, vol. 555, no. 7697, p. 487, 2018.
[20] T. Eo, Y. Jun, T. Kim, J. Jang, H.-J. Lee, and D. Hwang, "KIKI-net: cross-domain convolutional neural networks for reconstructing undersampled magnetic resonance images," *Magn Reson Med*, vol. 80, no. 5, pp. 2188-2201, 2018.
[21] J. Schlemper, J. Caballero, J. V. Hajnal, A. N. Price, and D. Rueckert, "A deep cascade of convolutional neural networks for dynamic MR image reconstruction," *IEEE Trans. Med. Imag.,* vol. 37, no. 2, pp. 491–503, 2018.
[22] T. M. Quan, T. Nguyen-Duc, and W.-K. Jeong, "Compressed sensing MRI reconstruction using a generative adversarial network with a cyclic loss," *IEEE Trans. Med. Imag.,* vol. 37, no. 6, pp. 1488–1497, 2018.
[23] Y. Yang, J. Sun, H. Li, and Z. Xu, "Deep ADMM-Net for Compressive Sensing MRI," presented at *the 30th Conference on Neural Information Processing Systems*, Barcelona, Spain, 2016.
[24] K. Hammernik, T. Klatzer, E. Kobler, et al. "Learning a Variational Network for Reconstruction of Accelerated MRI Data," *Magn Reson Med*, vol. 79, no. 6, pp. 3055-3071, 2018.
[25] C. Qin, J. V. Hajnal, D. Rueckert, J. Schlemper, J. Caballero, and A. N. Price, "Convolutional recurrent neural networks for dynamic MR image reconstruction," *IEEE Trans. Med. Imag.,* vol. 38, no. 1, pp. 280-290, 2019.
[26] S. Nassirpour, P. Chang, and A. Henning. "MultiNet PyGRAPPA: Multiple neural networks for reconstructing variable density GRAPPA (a 1H FID MRSI study)." *NeuroImage*, vol. 183, pp.336-345, 2018
[27] J. Cheng, H. Wang, L. Ying, and D. Liang, "Learning Primal Dual Network for Fast MR Imaging," *In the Proc. 27th Annual Meeting of ISMRM*, Montreal, QC, Canada, no. 4775, 2019.
[28] J. Yang, Y. Zhang, W. Yin, et al. "A Fast Alternating Direction Method for TVL1-L2 Signal Reconstruction From Partial Fourier Data," *IEEE JSTSP*, vol. 4, no. 2, pp. 288-297, 2010.
[29] H. Wang, L. Tam, E. Kopanoglu, D.C. Peters, R.T. Constable, G. Galiana. "Experimental O-space turbo spin echo imaging," *Magn Reson Med.* vol. 75, no. 4, pp. 1654-61, 2016
[30] H. Wang, L.K. Tam, E. Kopanoglu, D.C. Peters, R.T. Constable, G. Galiana. "O-space with high resolution readouts outperforms radial imaging," *Magn Reson Imaging*. vol. 37, pp. 107-115, 2017.